\documentstyle[preprint,aps]{revtex}
\input epsf
\begin{document}
\title{Phase Diagrams 
of $S=3/2, 2$ XXZ Spin Chains with Bond-Alternation}
\author{Atsuhiro Kitazawa$^{1,2}$
and Kiyohide Nomura$^{1}$}
\address{$^{1}$Department of Physics, Kyushu University, 
Fukuoka 812-12, Japan \\
$^{2}$Department of Physics, Tokyo Institute of Technology, 
Oh-okayama, Meguro-ku, Tokyo 152, Japan 
}
\date{\today}
\maketitle
\begin{abstract}
We study the phase diagram of $S=3/2$ and $S=2$ bond-alternating 
spin chains numerically. 
In previous papers, the phase diagram of $S=1$ XXZ spin chain with bond-alternation 
was shown to reflect the hidden $Z_{2}\times Z_{2}$ symmetry. 
But for the higher $S$ Heisenberg spin chain, 
the successive dimerization transition occurs, and for anisotropic spin chains 
the phase structure will be more colorful than the $S=1$ case. 
Using recently developed methods, 
we show directly that the phase structure of the anisotropic spin chains 
relates to the $Z_{2}\times Z_{2}$ symmetry. 
\end{abstract}

Haldane predicted that the antiferromagnetic Heisenberg spin 
chain has different behaviors between integer and half odd integer 
spins\cite{Haldane}.  
For half odd integer spin cases, the ground state 
has gapless excitations and spin correlations decay in the power law. 
The ground state properties for half odd spin Heisenberg chains 
are the same ones of the $S=1/2$ case. 
On the other hand for integer spin cases, 
the ground state is a singlet with finite 
excitation gap and the spin correlations decay exponentially. 
Another aspect for the Haldane gapped state was presented 
by Affleck {\it et.al.}\cite{AKLT}. 
They studied a special $S=1$ $SU(2)$ spin chain 
which has the property of the Haldane gap conjecture, 
and proposed valence bond solid (VBS) states. 
The $S=1$ VBS state has the hidden antiferromagnetic structure. 
Later den Nijs and Rommelse\cite{string} found the string order parameter 
as the order parameter for it.

Using a non-local unitary transformation, 
Kennedy and Tasaki\cite{Kennedy} showed that 
the broken symmetry of $S=1$ Haldane 
gap systems is a $Z_{2}\times Z_{2}$ type. 
Nearly four-fold degenerate states 
of the system with open boundary conditions 
reflect the discrete symmetry. 
For the higher $S$ cases, Oshikawa\cite{Oshikawa} showed that 
odd integer $S$ VBS states break the hidden $Z_{2}\times Z_{2}$ symmetry, 
while even integer $S$ VBS states 
do not violate the hidden $Z_{2}\times Z_{2}$ symmetry, 
and considered the generalized string order parameters. 
He pointed out another possibilities 
for the broken symmetry of higher $S$ cases. 

In our previous papers with Okamoto\cite{KNO,jpsj}, 
we studied the $S=1$ XXZ spin chains with the  bond-alternation, 
\begin{equation}
  H = \sum_{j=1}^{N}\left[1-(-1)^{j}\delta\right]
  \left[ S^{x}_{j}S^{x}_{j+1}+S^{y}_{j}S^{y}_{j+1}+\Delta S^{z}_{j}S^{z}_{j+1}
  \right]
\label{hamiltonian}
\end{equation}
and we obtained the phase diagram reflecting the hidden 
$Z_{2}\times Z_{2}$ symmetry. 
Comparing with the phase structure 
of the quantum Ashkin-Teller model\cite{Kohmoto}
which has a $Z_{2}\times Z_{2}$ symmetry explicitly, 
the possible phase diagram consists of 
the two dimensional(2D) Ising, the Gaussian, 
and the Berezinskii-Kosterlitz-Thouless(BKT) transitions. 
Combining the quantum Ashkin-Teller picture with the generalized hidden 
$Z_{2}\times  Z_{2}$ symmetry, 
we predicted that for $-1<\Delta$, $-1<\delta<1$, 
there are $2S+1$ 2D Ising, $2S$ Gaussian, and 
$2S+1$ BKT critical lines. 
Especially the Gaussian lines should separate 
several VBS states, as was argued by 
Guo {\it et.al.}\cite{Guo} and Oshikawa\cite{Oshikawa}. 
According to them, 
the VBS state with the periodic boundary condition can be written by
\begin{eqnarray}
  \lefteqn{ |S,M,\mbox{PBC}\rangle } \nonumber \\
    &&= (a^{\dagger}_{N}b^{\dagger}_{1}-b^{\dagger}_{N}a^{\dagger}_{1})^{S-M}
  \times
\nonumber \\
  && \prod_{j=1}^{N/2-1}
  (a^{\dagger}_{2j-1}b^{\dagger}_{2j}
    -b^{\dagger}_{2j-1}a^{\dagger}_{2j})^{S+M} 
  (a^{\dagger}_{2j}b^{\dagger}_{2j+1}-b^{\dagger}_{2j}a^{\dagger}_{2j+1})^{S-M}
  \nonumber \\
  &&\times (a^{\dagger}_{N-1}b^{\dagger}_{N}
    -b^{\dagger}_{N-1}a^{\dagger}_{N})^{S+M}|0\rangle,
\label{stt}
\end{eqnarray}
where we describe the spin state by the Schwinger bosons, that is, 
$a_{j}^{\dagger}$($b_{j}^{\dagger}$) creates the $S=1/2$ $\uparrow$ ($\downarrow$) 
state at the $j$-th site, and $M$ is an integer for integer $S$, or 
a half integer for half integer $S$, 
which changes from $-S$ to $S$ if we vary $\delta$ from $-1$ to $1$. 
Hereafter we denote this VBS state as $M$-VBS state. 
Oshikawa\cite{Oshikawa} showed that for integer $S$ case, 
the hidden $Z_{2}\times Z_{2}$ symmetry is broken in the $M$-VBS state 
when $S-M$ is an odd integer.
For the isotropic case ($\Delta=1$), the $S=1$ case is known that the transition 
between the Haldane and the dimer phases occurs about 
$\delta=0.26$\cite{Kato,Yamamoto,Totsuka,Yamamoto97,jpsj}. 
For the $S=3/2$ case, Yajima and Takahashi\cite{Yajima} 
studied with the density matrix 
renormalization group method and evaluated the transition point between 
$M=1/2$ and $M=3/2$ phase as $\delta_{1/2-3/2}=0.42\pm0.02$. 
Yamamoto\cite{Yamamoto97} obatained consistent results as 
$\delta_{1/2-3/2}=0.43\pm0.01$ by quantum Monte Calro calculation. 
For the $S=2$ case, Yamanaka, Oshikawa, and Miyashita\cite{Yamanaka} 
evaluated the transition 
point by quantum Monte Carlo calculation. 
They obtained that the transition point between $M=0$ and $M=1$ phases 
is in the region $0.05 < \delta_{0-1} < 0.3$, 
and the transition point between $M=1$ and $M=2$ is in the 
region $0.5 < \delta_{1-2} < 0.6$. 
Yamamoto\cite{Yamamoto97} evaluated the transition point 
of $S=2$ case more accurately 
as $\delta_{0-1} = 0.18\pm 0.01$, $\delta_{1-2}=0.545\pm 0.005$. 

In this paper, we show the phase diagram of the model(\ref{hamiltonian}) 
for $S=3/2$ and $S=2$ cases with the exact diagonalization method 
for finite size systems. 
In the following, we mainly argue the successive dimerization 
of the model(\ref{hamiltonian}). 

We use the following sine-Gordon Euclidean action 
as an effective model\cite{Schulz} of the Hamiltonian(\ref{hamiltonian})
\begin{eqnarray}
  S &=& \frac{1}{2\pi}\int vd\tau dx\frac{1}{K}
    \left[ \left(\frac{\partial\phi}{v\partial \tau}\right)
         + \left(\frac{\partial\phi}{\partial x}\right)\right]
\label{SG} \\
    &&+\frac{y_{1}}{2\pi a^{2}}\int vd\tau dx\cos\sqrt{2}\phi
     +\frac{y_{2}}{2\pi a^{2}}\int vd\tau dx\cos\sqrt{8}\phi,
\nonumber
\end{eqnarray}
where $a$ is a lattice constat and $v$ is the sound velocity. 
The dual field $\theta$ is defined as 
\begin{eqnarray}
  \frac{\partial}{v\partial\tau}\phi(\tau,x)
    &=&-\frac{\partial}{\partial x}(iK\theta(\tau,x)),\hspace{5mm}
\nonumber \\
  \frac{\partial}{\partial x}\phi(\tau,x)
    &=&\frac{\partial}{v\partial \tau}(iK\theta(\tau,x)).
\nonumber
\end{eqnarray}
We make the identification $\phi\equiv\phi+\sqrt{2}\pi$, 
$\theta\equiv\theta+\sqrt{2}\pi$. 
For the free field theory, the primary operators 
$\exp(in\sqrt{2}\theta+im\sqrt{2}\phi)$ have the scaling dimension 
$x_{n,m}=n^{2}/2K+m^{2}K/2$ and the spin $s_{n,m}=nm$ (where integer 
variables $n$ and $m$ are electric and magnetic charges in the Coulomb gas 
picture). 
The second term of eq.(\ref{SG}) is the mass term for the 
Haldane gap systems\cite{Schulz,Affleck86}. 
(According to Affleck\cite{Affleck86}, we have $y_{1}\propto\cos\Theta/2$, 
where $\Theta$ is the topological angle.)

After scaling $a\rightarrow ae^{dl}$, we obtain the following 1-loop 
renormalization group equations
\begin{eqnarray}
  \frac{d}{dl}\frac{1}{K} &=& \frac{1}{8}y_{1}^{2}+\frac{1}{2}y_{2}^{2},
  \nonumber \\
  \frac{dy_{1}}{dl} &=& \left( 2-\frac{K}{2}\right)y_{1} 
    -\frac{1}{2}y_{1}y_{2},
  \\
  \frac{dy_{2}}{dl} &=& \left( 2-2K\right)y_{2} -\frac{1}{4}y_{1}^{2}.
  \nonumber
\end{eqnarray}
From these equations we can neglect the third term of eq.(\ref{SG}) for $K>1$. 
The third term is important for the 2-D Ising transition 
and the level-1 $SU(2)$ WZW point, 
but not for the BKT and the Gaussian transitions. 
Neglecting $y_{2}$ we obtain the following Kosterlitz's recursion relation
\begin{equation}
  \frac{d}{dl}\frac{1}{K} = \frac{1}{8}y_{1}^{2},
  \hspace{5mm}
  \frac{dy_{1}}{dl} = \left( 2-\frac{K}{2}\right)y_{1}.
\label{KTeq}
\end{equation}
Near $K=4$, there occurs the BKT transition, and for $4>K>1$ the Gaussian 
transition occurs at $y_{1}=0$. 

\begin{figure}
\begin{center}
\epsfxsize=4.0in 
\leavevmode\epsfbox{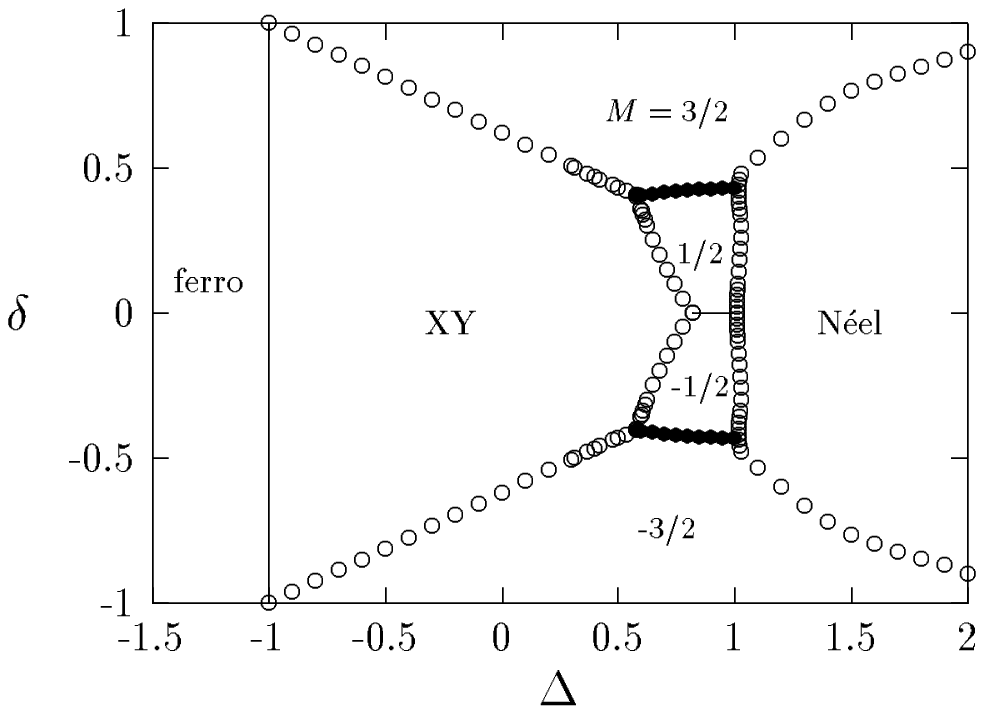}
\end{center}
\vspace{3mm}
\caption{Phase diagram of the $S=3/2$ model. 
The solid circles and the horizontal line separate 
$M=-3/2,-1/2,1/2,3/2$ VBS phases. Transitions between the XY and the VBS 
phases are of the BKT type. Transitions between the N\'eel and the VBS 
phases are of the 2D Ising type.}
\label{pdgrm}
\end{figure}

\begin{figure}[h]
\begin{center}
\epsfxsize=4.0in 
\leavevmode\epsfbox{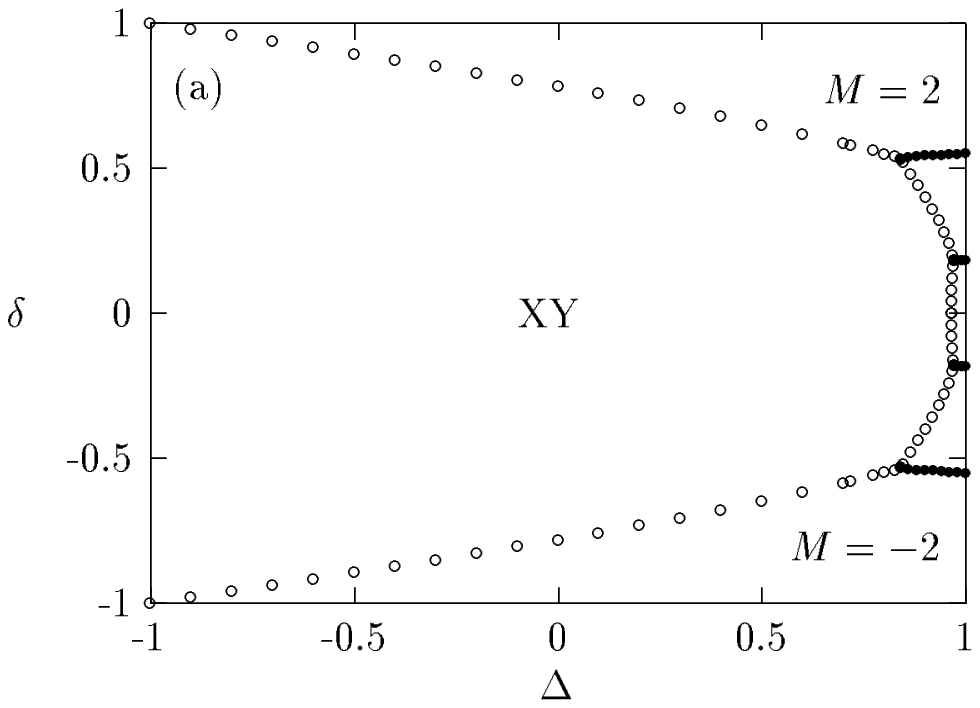}
\vspace{5mm}

\epsfxsize=2.1in 
\leavevmode\epsfbox{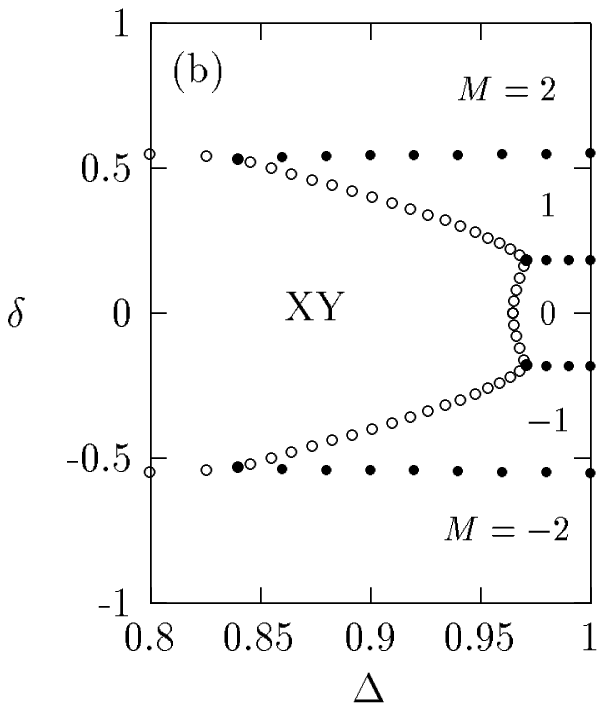}
\end{center}
\vspace{3mm}
\caption{Phase diagram of the $S=2$ model in the region (a):$-1<\Delta<1$, 
and (b):$0.8<\Delta<1$. Solid circles separate $M=-2,-1,0,1,2$ VBS phases.}
\label{pdgrm2}
\end{figure}

The obtained phase diagrams are in Fig.\ref{pdgrm} for $S=3/2$ case 
and in Fig.\ref{pdgrm2} for $S=2$ case 
using the method in \cite{Kitazawa,Nomura}. 
The topology of the phase diagrams is consistent 
with the prediction as is described above. 

At this stage, we will explain the method to determine the Gaussian line 
and the relation to the $Z_{2}\times Z_{2}$ symmetry. 
Considering the hidden $Z_{2}\times Z_{2}$ symmetry, 
the successive dimerized transitions are of the Gaussian type. 
The Hamiltonian is invariant under the 
space inversion $P$:$j\rightarrow N-1+j$ and 
the spin reversal $T$:$a^{\dagger}\leftrightarrow b^{\dagger}$ and 
all states(\ref{stt}) have the same eigenvalues 
of $P$ and $T$ ($P=T=(-1)^{SN}$, independent of $M$).
For the transition from $M$ phase to the $M+1$ one, 
the ground state is singlet 
in both phases, so we can not simply apply 
the phenomenological renormalization group technique. 
But with the twisted boundary condition
\[
  S^{x}_{N+1}\pm iS^{y}_{N+1}= -(S^{x}_{1}\pm iS^{y}_{1}), \hspace{5mm}
  S^{z}_{N+1} = S^{z}_{1},
\]
the $M$-VBS state(\ref{stt}) changes to\cite{jpsj}
\begin{eqnarray}
 \lefteqn{ |S,M,\mbox{TBC}\rangle } \nonumber \\
    &&= (a^{\dagger}_{N}b^{\dagger}_{1}+b^{\dagger}_{N}a^{\dagger}_{1})^{S-M} 
  \times  
\nonumber \\
  &&  \prod_{j=1}^{N/2-1}(a^{\dagger}_{2j-1}b^{\dagger}_{2j}
  -b^{\dagger}_{2j-1}a^{\dagger}_{2j})^{S+M}
  (a^{\dagger}_{2j}b^{\dagger}_{2j+1}-b^{\dagger}_{2j}a^{\dagger}_{2j+1})^{S-M}
  \nonumber \\
  &&\times (a^{\dagger}_{N-1}b^{\dagger}_{N}
  -b^{\dagger}_{N-1}a^{\dagger}_{N})^{S+M}|0\rangle.
\label{dimerstate}
\end{eqnarray}
From this equation, the $M$-VBS state has $P=T=(-1)^{SN-S+M}$ and 
the $M+1$ state has $P=T=(-1)^{SN-S+M+1}$, so the two states have 
the different symmetries, and the energy levels cross at the transition point. 

In the sine-Gordon Language, the effect of the twisted boundary condition 
is to shift the magnetic charge $m$ as $m\rightarrow m+1/2$\cite{Alcaraz}, 
so only the half integer magnetic charge operators exist. 
Then we have the following size dependence of the scaling dimensions 
of the $m=\pm1/2$ operators 
$\sqrt{2}\cos(\phi/\sqrt{2})$ [$x_{0,1/2}^{e}$] and 
$\sqrt{2}\sin(\phi/\sqrt{2})$ [$x_{0,1/2}^{o}$]\cite{Kitazawa},
\begin{eqnarray}
  x_{0,1/2}^{e}(L) &=& \frac{L(E^{e}_{L,\pi}-E^{e}_{L,0})}{2\pi v} 
    = \frac{K}{8} + \frac{y_{1}(L)}{2} +\cdots,
  \nonumber \\
  x_{0,1/2}^{o}(L) &=& \frac{L(E^{o}_{L,\pi}-E^{e}_{L,0})}{2\pi v} 
    = \frac{K}{8} - \frac{y_{1}(L)}{2} +\cdots,
  \nonumber
\end{eqnarray}
where $L$ is the system size, 
$y_{1}$ is a renormalized function of $L$ as 
\[
  y_{1}(L) = y_{1}\left(\frac{2\pi}{L}\right)^{K/2 -2},
\]
$E_{L,0}^{e}$ is the ground state energy of the periodic boundary 
conditions, and we denote $E_{L,\pi}^{e,o}$ as the corresponding 
energy to $\sqrt{2}\cos(\phi/\sqrt{2})$, 
$\sqrt{2}\sin(\phi/\sqrt{2})$ (e,o mean the parity $P=1$, $-1$). 
In this method, the correction from the third term of eq.(\ref{SG}) 
does not affect since this method uses 
the duality of the quantum Ashkin-Teller model\cite{jpsj}. 
Note that Kolb\cite{Kolb} studied the XXZ spin chain with 
the twisted boundary condition 
$  S^{\pm}_{N+1}= e^{\pm i\Phi}S^{\pm}_{1}$, 
$S^{z}_{N+1} = S^{z}_{1}$ and showed that for the half integer spin model 
the ground states are two-fold degenerate at $\Phi=\pi$, 
since for the half integer $S$ XXZ spin chains, 
the second term in eq.(\ref{SG}) is absent ($y_{1}=0$).

\begin{figure}
\begin{center}
\epsfxsize=3in 
\leavevmode\epsfbox{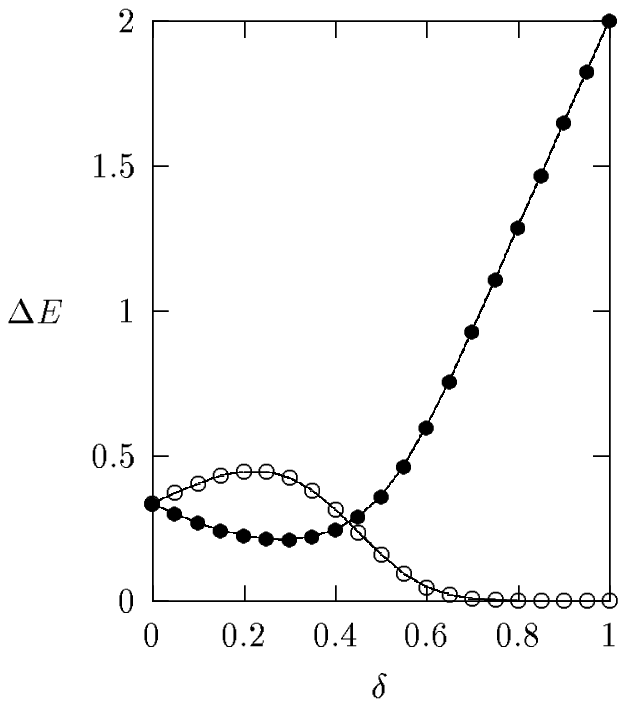}
\end{center}
\vspace{3mm}
\caption{$E^{e}_{N,\pi} - E^{e}_{N,0}$ ($P=T=1$) ($\circ$) 
and $E^{o}_{N,\pi} - E^{e}_{N,0}$ ($P=T=-1$) ($\bullet$) 
of $S=3/2$, $\Delta=1$, $N=12$  
systems.}
\label{crss3}
\end{figure}

\begin{figure}
\begin{center}
\epsfxsize=3.5in
\leavevmode\epsfbox{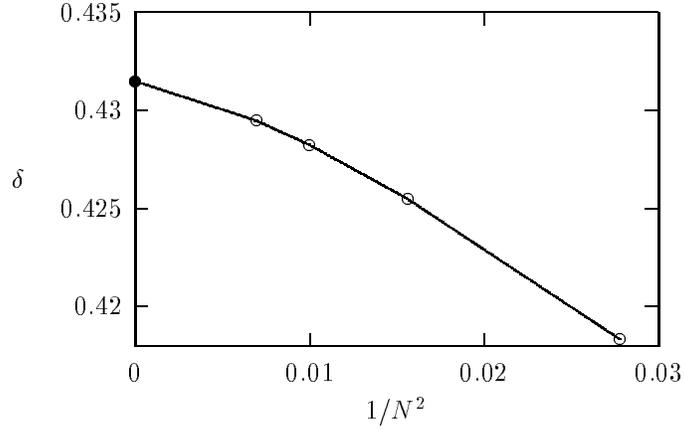}
\end{center}
\caption{Size dependence of the crossing points ($\delta_{1/2-3/2}$) 
for $S=3/2$ and  $\Delta=1$.}
\label{point}
\end{figure}

For $S=3/2$ case, we study the $N=6,8,10,12$ systems. 
In Fig. \ref{crss3}, we show the two low lying energies 
of the subspace $\sum S^{z}=0$ 
with the twisted boundary condition for $\Delta=1$, $N=12$. 
For small $\delta(>0)$ the lowest state has $P=T=1$, 
and for $1>\delta>\delta_{1/2-3/2}$ the lowest state has $P=T=-1$, 
as expected from eq.(\ref{dimerstate}). 
Fig. \ref{point} shows the size dependence of the crossing point. 
We extrapolated the crossing point as polynomials of $1/N^{2}$\cite{jpsj} 
and the evaluated value 
is $\delta_{1/2-3/2}=0.4315$, which is consistent 
with the previously obtained value \cite{Yajima,Yamamoto97}. 
For the BKT transitions between the XY and the VBS phases are determined 
by the similar method in ref \cite{Nomura} (see also\cite{KNO,jpsj}). 
The evaluated multi-critical points of the XY, $M$-VBS, and $M+1$ VBS phases 
are $(\Delta_{M}, \delta_{M})=(0.8188,0)$, 
$(0.5766,\pm 0.4052)$. 
For the transitions between the N\'eel and the VBS phases (2D Ising type), 
we use the phenomenological renormalization group method.

\begin{figure}
\begin{center}
\epsfxsize=3in 
\leavevmode\epsfbox{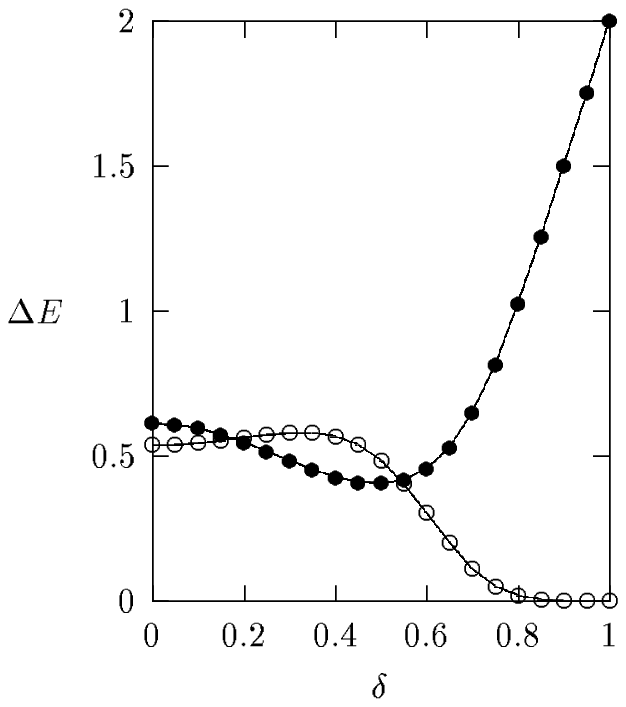}
\end{center}
\vspace{3mm}
\caption{$E^{e}_{N,\pi} - E^{e}_{N,0}$ ($P=T=1$) ($\circ$) 
and $E^{o}_{N,\pi} - E^{e}_{N,0}$ ($P=T=-1$) ($\bullet$) 
of $S=2$, $\Delta=1$, $N=10$  
systems.}
\label{crss4}
\end{figure}

\begin{figure}
\begin{center}
\epsfxsize=3.5in
\leavevmode\epsfbox{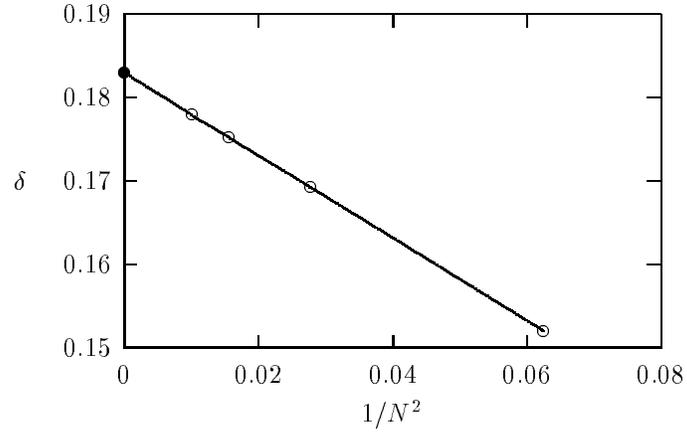}
\end{center}
\caption{Size dependence of the crossing points ($\delta_{0-1}$) 
for $S=2$ and  $\Delta=1$.}
\label{pointg1}
\end{figure}

For $S=2$ case, we calculate the energy eigenvalues for $N=4,6,8,10$. 
In Fig. \ref{crss4}, we show the two low lying energies of the subspace $\sum S^{z}=0$ 
with the twisted boundary condition for $\Delta=1$, $N=10$. 
The level crossing occurs two times for $0<\delta<1$. 
For $0<\delta<\delta_{0-1}$ the lowest state has $P=T=1$ 
($S=2$ Haldane phase, $M=0$), 
for $\delta_{0-1}<\delta<\delta_{1-2}$ the lowest state has $P=T=-1$ ($M=1$), 
and for $\delta_{1-2}<\delta<1$ the lowest state has $P=T=1$ ($M=2$), 
as expected from eq.(\ref{dimerstate}). 
The hidden $Z_{2}\times Z_{2}$ symmetry is broken for $M=\pm 1$ phase. 
Fig. \ref{pointg1} shows the size dependence of the crossing point 
($\delta_{0-1}$). 
The extrapolated values of the transition points are $\delta_{0-1}=0.1830$ 
(between $M=0$ and $M=1$ VBS phases), 
and $\delta_{1-2}=0.5505$ (between $M=1$ and $M=2$ VBS phases). 
These values are consistent with the results of Yamamoto\cite{Yamamoto97}.
For $\delta=0$, the transition point between the XY and the $S=2$ 
Haldane phases is $\Delta_{c}=0.9650$. 
The evaluated multi-critical points between the XY and the VBS phases 
are $(\Delta_{M}, \delta_{M})=(0.9708,\pm 0.1823)$ and  
$(0.8401,\pm 0.5325)$. 

In this paper, we showed the phase diagrams of the $S=3/2, 2$ bond-alternating 
XXZ spin chain. On the base of the hidden $Z_{2}\times Z_{2}$ symmetry, 
we can see the successive dimerization numerically using the 
twisted boundary conditions. 
The two lowest excitations in the twisted boundary condition 
correspond to the order 
parameter and its dual disorder parameter (with respect to 
the $Z_{2}\times Z_{2}$ symmetry)\cite{jpsj}, 
so their crossing point is the self-dual point (in the Ashkin-Teller language) 
or the Gaussian critical point. 
Therefore, we can determine the critical point more precisely than the method 
using the string order parameter only\cite{Totsuka}, 
especially near the BKT multicritical point. 

We acknowledge H. Shiba, K. Okamoto, and S. Yamamoto. 
This work is partly supported by Grant-in-Aid for Scientific Research (C) 
No.09740308 from the Ministry of Education, Science and Culture, Japan.
A.K. is supported by JSPS Research Fellowships for Young Scientists. 
The numerical calculation was done 
using the facilities of the Super Computer Center, 
Institute for Solid State Physics, University of Tokyo.

\clearpage

\clearpage

\end{document}